# Cortical lesions, central vein sign, and paramagnetic rim lesions in multiple sclerosis: emerging machine learning techniques and future avenues

Francesco La Rosa[1,2], Maxence Wynen[3,2,4], Omar Al-Louzi[5,6], Erin S Beck[5,7], Till Huelnhagen[1,4,8], Pietro Maggi[9,10], Jean-Philippe Thiran[1,2,4], Tobias Kober[1,4,8], Russell T Shinohara[11,12,13], Pascal Sati[6], Daniel S Reich[5], Cristina Granziera[14,15], Martina Absinta[16,17], Meritxell Bach Cuadra[2,4]

[1]Signal Processing Laboratory (LTS5), Ecole Polytechnique Fédérale de Lausanne (EPFL), Lausanne, Switzerland.
[2]CIBM Center for Biomedical Imaging, Switzerland.
[3]ICTeam, UCLouvain, Louvain-la-Neuve, Belgium.
[4]Radiology Department, Lausanne University and University Hospital, Switzerland.
[5]Translational Neuroradiology Section, National Institute of Neurological Disorders and Stroke, National Institutes of Health, Bethesda, MD 20892, USA.
[6]Department of Neurology, Cedars-Sinai Medical Center, Los Angeles, CA 90048, USA.
[7]Department of Neurology, Icahn School of Medicine at Mount Sinai, New York, NY, 10029, USA.
[8]Advanced Clinical Imaging Technology, Siemens Healthcare AG, Lausanne, Switzerland.
[9]Department of Neurology, Cliniques universitaires Saint-Luc, Université catholique de Louvain, Brussels, Belgium.
[10]Department of Neurology, CHUV, Lausanne, Switzerland
[11]Center for Biomedical Image Computing and Analysis (CBICA), Department of Radiology, University of Pennsylvania, Philadelphia, PA,19104.
[12]Penn Statistics in Imaging and Visualization Endeavor (PennSIVE), Center for Clinical Epidemiology and Biostatistics, University of Pennsylvania, Philadelphia, PA, 19104.
[13]Department of Biostatistics, Epidemiology, and Informatics, Perelman School of Medicine, University of Pennsylvania, Philadelphia, PA, USA.
[14]Translational Imaging in Neurology (ThINk) Basel, Department of Biomedical Engineering, Faculty of Medicine, University Hospital Basel and University of Basel, Basel, Switzerland.
[15]Neurologic Clinic and Policlinic, MS Center and Research Center for Clinical Neuroimmunology.
[16]IRCCS San Raffaele Hospital and Vita-Salute San Raffaele University, Milan, Italy.
[17]Department of Neurology, Johns Hopkins University School of Medicine, Baltimore, MD, USA.

Corresponding author: Francesco La Rosa
Signal Processing Laboratory (LTS5). EPFL-STI-IEL-LTS5 Station 11. CH-1015 Lausanne, Switzerland. Telephone: +41 21 693 78 89. E-mail: francesco.larosa@epfl.ch.

# Abstract

The current multiple sclerosis (MS) diagnostic criteria lack specificity, and this may lead to misdiagnosis, which remains an issue in present-day clinical practice. In addition, conventional biomarkers only moderately correlate with MS disease progression. Recently, advanced MS lesional imaging biomarkers such as cortical lesions (CL), the central vein sign (CVS), and paramagnetic rim lesions (PRL), visible in specialized magnetic resonance imaging (MRI) sequences, have shown higher specificity in differential diagnosis. Moreover, studies have shown that CL and PRL are potential prognostic biomarkers, the former correlating with cognitive impairments and the latter with early disability progression. As machine learning-based methods have achieved extraordinary performance in the assessment of conventional imaging biomarkers, such as white matter lesion segmentation, several automated or semi-automated methods have been proposed for CL, CVS, and PRL as well. In the present review, we first introduce these advanced MS imaging biomarkers and their imaging methods. Subsequently, we describe the corresponding machine learning-based methods that were used to tackle these clinical questions, putting them into context with respect to the challenges they are still facing, including non-standardized MRI protocols, limited datasets, and moderate inter-rater variability. We conclude by presenting the current limitations that prevent their broader deployment and suggesting future research directions.

**Abbreviations:** Multiple sclerosis (MS), (MRI), cortical lesions (CL), white matter lesions (WML), convolutional neural network (CNN), paramagnetic rim lesions (PRL), central vein sign (CVS), machine learning (ML), deep learning (DL), white matter (WM), gray matter (GM).

# 1   Introduction

Multiple sclerosis (MS) is a chronic inflammatory disease and a common cause of neurological disability in young adults[1]. Its hallmark is demyelinated white matter lesions (WML) forming in the central nervous system[1]. These lesions are assessed in-vivo with magnetic resonance imaging (MRI), which is the imaging technique of choice to diagnose MS and monitor the disease over time[2]. The current MRI diagnostic criteria (McDonald criteria) are based on the dissemination in space and time of such lesions[3]. Moreover, the quantification of the total lesion volume is important to determine ongoing disease activity and monitor treatment effect over time[4]. Recommended MRI sequences include dark-blood T2-weighted contrasts, such as fluid-attenuated inversion recovery (FLAIR), and T1-weighted contrasts, such as magnetization prepared rapid gradient-echo (MPRAGE)[5]. At common clinical magnetic fields (1.5T and 3T), the use of gadolinium-based contrast agents is useful to evaluate patients suspected of MS and monitor disease activity causing breakdown of the blood-brain barrier[6].

As the manual detection of WML is time-consuming and prone to inter-rater variability[7], a myriad of automated or semi-automated approaches have been developed to facilitate this task[8], representing some of the earliest uses of machine learning techniques applied to MRI. These methods were initially based primarily on MRI intensity features and probabilistic atlases[8], whereas more recently, the vast majority use deep learning (DL) approaches[9], the latter without prior feature extraction. Substantial effort is now being made towards reproducibility of the results and open science[10]. Several grand challenges have been organized, with the respective training datasets released publicly in order to provide a fair evaluation and benchmarking of automated techniques[11–13]. In the latest contests, DL-based methods have achieved the best performance, approaching or sometimes even outperforming human readers[11,13]. WML segmentation methods have been reviewed relatively recently[9,14]; the present review thus focuses on automated techniques tailored for advanced imaging biomarkers specific to MS that have the potential to improve MS diagnosis and prognosis.

One major drawback of the current MS diagnostic criteria is their lack of specificity, as they were proposed to identify patients with a high likelihood of MS rather than distinguish MS from other conditions[3]. The lack of specificity of these criteria may lead to misdiagnosis, which remains a persistent problem of MS[15]. Multi-center studies have shown a misdiagnosis rate of 18%[16], often associated with atypical clinical or imaging findings. Improving the diagnostic specificity would prevent harmful consequences for patients[17] and allow clinicians to prescribe the appropriate treatment earlier. In addition, although clinical relapses are often associated with the appearance of new WML, the overall WML burden, which is the most common MRI biomarker examined in clinical routine, is only moderately correlated with disability and poorly predicts transition to progressive disease[18]. For all these reasons, there is a need for additional biomarkers that are highly specific to MS or correlate with disease progression.

Recently, advances in MR technology, such as the development of specialized sequences, acceleration of protocols, and the proliferation of ultra-high field MRI, have allowed the imaging of novel and pathologically specific MS biomarkers[19,20]. These include cortical lesions (CL), the central vein sign (CVS), and paramagnetic rim lesions (PRL). On the one hand, studies have shown that CL and PRL are potential prognostic biomarkers: CL are associated with cognitive impairments, while patients with PRL experience an earlier progression in disability[21,22]. On the other hand, the CVS and PRL have proven to be effective for differentiating MS from mimicking diseases[23–26]. All three biomarkers, however, require dedicated MRI sequences at high (3T) or ultra-high (7T) magnetic fields, and experienced raters for their manual assessment, which can be very time-consuming. As done in the past for WML, various automated or semi-automated methods, mostly based on machine learning (ML), have been developed to facilitate the advanced biomarkers' assessment (see Table 1). Compared to their WML counterparts, however, they face additional challenges, including non-standardized imaging protocols, moderate inter-rater variability when determining ground truth annotations, and smaller datasets. Automated assessment could improve standardization and facilitate large-scale assessment in clinical routine of the aforementioned advanced biomarkers, with clear benefits in terms of MS diagnosis and prognosis.

In this review, we first briefly describe these novel brain imaging biomarkers and their imaging requirements and then focus on image processing techniques tailored for their automated segmentation and classification. We conclude with a discussion on current limitations and future lines of research to boost the development of ML approaches in this area and encourage their adoption in MS research and clinical settings.

# 2   Advanced MRI biomarkers in multiple sclerosis

In this section, we present a brief description of CL, CVS, and PRL, and their respective imaging protocols. In addition to the CVS and PRL, which have emerged as promising MS biomarkers in recent years, we also included CL which, although included in the MS diagnostic criteria, are not yet commonly analyzed in clinical practice.

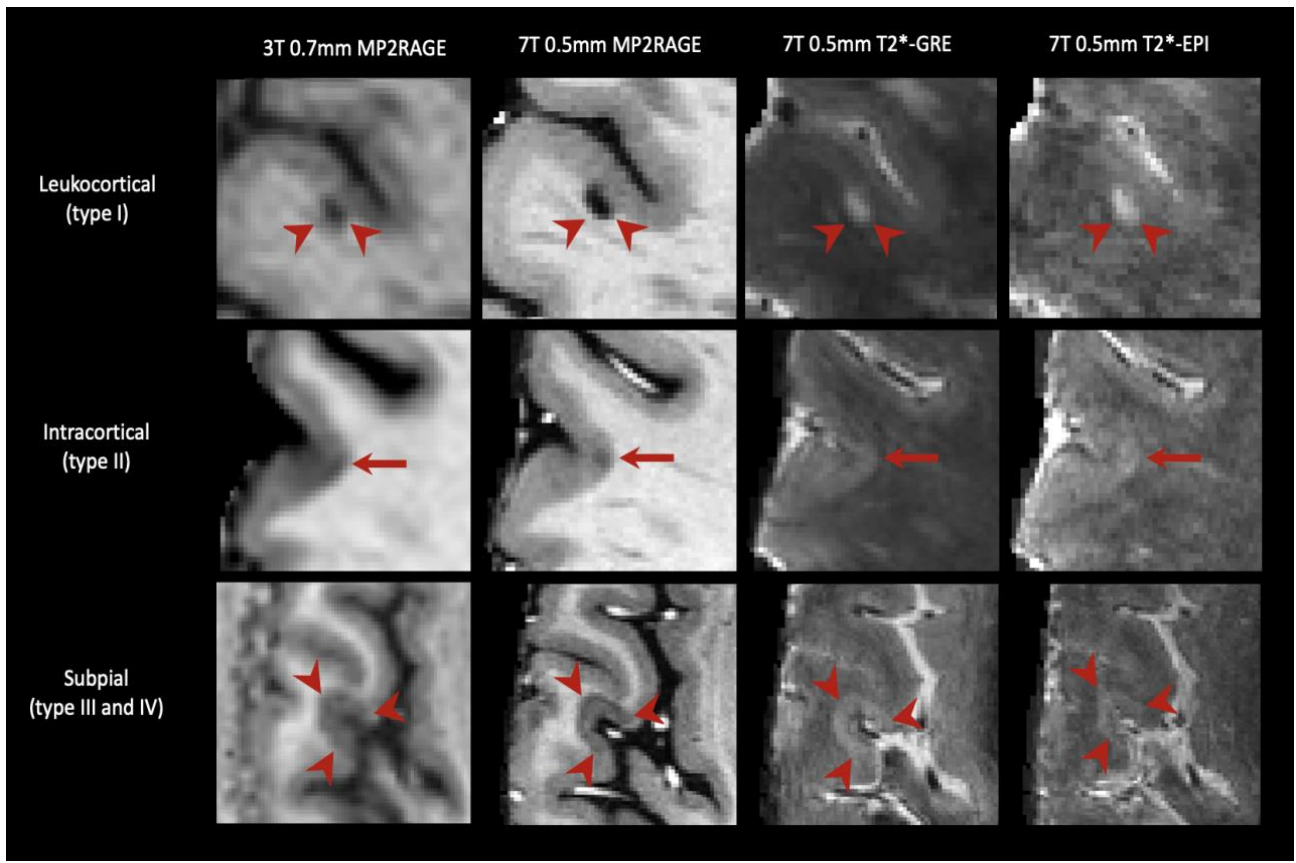

*Figure 1. Cortical lesions, including leukocortical, intracortical, and subpial subtypes, are seen better at 7T due to higher signal-to-noise ratios, allowing higher resolution scans, and increased susceptibility effects. The 7T 0.5mm MP2RAGE image shown was obtained as the average of 4 acquisitions.*

**Cortical lesions (CL) -** Cortical lesions are a type of MS lesions that involve, at least partially, the cortex and have been classified into three main categories[21] (see Figure 1): leukocortical lesions are located at the interface between WM and gray matter (GM) (type I), intracortical lesions are purely in the cortex and do not reach the pial surface (type II), and subpial lesions touch the subpial surface of the brain (type III) and sometimes extend all the way to the white matter (type IV). Cortical demyelination in MS has long been recognized in pathology studies, but only in the last two decades dedicated sequences on high- and ultra-high field scanners have provided in-vivo evidence of cortical damage[21]. Cortical lesions have raised great interest for several reasons. First, they have been observed in the early stages of the disease and in all of the major MS phenotypes[27]. Second, they are associated with disability[28–30] and in some studies, their number was associated with cognitive disability more strongly than WML[28,31]. Third, longitudinal studies have linked them with disease progression[32–35]. Fourth, subpial cortical demyelination is highly specific to MS[36]; CL have been observed in patients with radiologically isolated syndrome[37], but not in patients with neuromyelitis optica[38]. Since 2017, cortical lesions have been included in the MS diagnostic criteria[3], but their visualization on routine MRI sequences remains difficult. For instance, a postmortem study showed that 3D FLAIR at 3T could detect about 41% of leukocortical lesions and only 5% of the purely cortical ones[39]. This supports the need for specialized MRI techniques such as the phase-sensitive inversion recovery (PSIR), double inversion recovery (DIR), and magnetization-prepared 2 rapid gradient echoes (MP2RAGE)[6]. However, these sequences are still relatively insensitive to CL at 1.5T and 3T[40,41]. Ultra-high field MRI, with its higher signal-to-noise ratio and increased susceptibility effects, has proven to be a powerful tool for increasing the sensitivity to CL, especially for intracortical and subpial lesions[42]. Even with the most sensitive methods, however, CL are small and often subtle, making manual segmentation extremely time consuming and subject to relatively low inter-rater reliability[28,43].

**Central vein sign (CVS) -** Recently, studies have suggested that an MRI-detectable central vein inside MS lesions might be evidence of pathological processes specific to MS (see Figure 2)[44,45]. This marker, referred to as the "central vein sign," has gained attention in recent years, as it could help to differentiate MS from mimicking diseases[24,46]. Small cerebral veins can be detected with susceptibility-based MRI sequences, taking advantage of the magnetic properties of venous blood that is rich in deoxyhemoglobin[47,48]. To obtain the best detection sensitivity for the CVS, optimized MRI acquisitions have been proposed (T2*-weighted acquired with 3D-segmented echo-planar-imaging or T2*w 3D-EPI[49], combined T2-FLAIR and T2*, also called FLAIR*[50]). These sequences were shown to provide superior CVS detection compared to clinical acquisitions at 1.5T and 3T[51,52]. Single-center and multi-center retrospective studies imaging patients with clinically established diagnoses have demonstrated a significantly higher proportion of CVS-positive white matter lesions (%CVS+) in MS (mean pooled incidence: 79%, 95% CI: 68–87%)[52] as compared to other neurological disorders mimicking MS (mean pooled incidence: 38%, 95% CI: 18–63%)[52] such as cerebral small vessel disease[53], neuromyelitis optica spectrum disorder (NMOSD)[54], inflammatory vasculopathies[44], and migraine[45]. To distinguish MS from other neurological conditions, different CVS-based criteria have been proposed to date, some based on the percentage of perivenular lesions (from 35% to 60%) and others simply on the CVS lesion count (3-lesion or 6-lesion rule)[44,55–57]. From a diagnostic perspective, retrospective studies have shown excellent diagnostic discrimination by applying the '40% rule'[55] with sensitivity = 91% [95% CI, 82%-97%] and specificity = 96% [95% CI, 88%-100%])[51]. However, applying percentage-based criteria requires manual exclusion of lesions that are confluent or have multiple or eccentric veins, and performing the CVS evaluation on all the remaining lesions present in patients' brains, which is a time-consuming process difficult to accomplish in clinical practice.

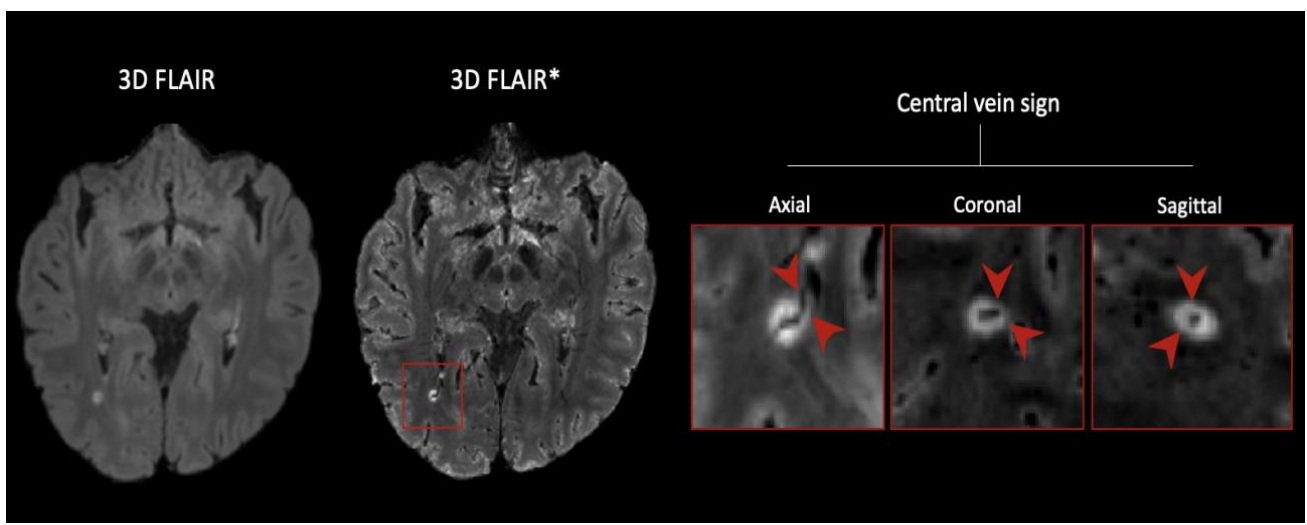

Figure 2. A central vein running through a lesion visible in the three planes (zoomed-in boxes) in a 3D FLAIR* obtained at 3T. Resampling was applied to the magnified images for visualization purposes.

**Paramagnetic rim lesions (PRL) -** Recent pathological studies have demonstrated that about 30% of chronic demyelinated lesions are pathologically characterized by perilesional accumulation of iron-laden microglia and macrophages, showing evidence of smouldering demyelination and axonal loss around an inactive hypocellular core (see Figure 3)[58,59]. This type of MS lesion has been defined as "chronic active/smouldering lesions". Due to their peripheral paramagnetic iron rim, these lesions can be depicted on in-vivo susceptibility-based MRI techniques (T2*-weighted magnitude, phase images, and quantitative susceptibility mapping, QSM) at both 3T and 7T[60,61], and are therefore termed "paramagnetic rim lesions" (PRL). Recent in vivo studies have shown that about 50% of relapsing and about 60% of progressive MS patients have at least one PRL[22,26]. Of clinical relevance, PRL accrual has been recently linked to a more aggressive disease course and disability accumulation at a younger age and/or shorter disease duration[22]. Reasons for such association directly rely on a few typical features of these lesions: PRL are destructive[61,62], they do not remyelinate[61], and they can expand over time[22] demyelinating the surrounding tissue and injuring axons, as corroborated by the elevation of serum neurofilament light chain in patients with PRL who are not forming new white matter lesions[63]. The recent discovery that the paramagnetic rim can disappear over a median of 7 years[64] holds promise regarding its potential use as an outcome measure in clinical trials designed to halt the chronic inflammation at the lesion edge.

In addition to their prognostic role, PRL appear specific to MS, as they have been rarely detected in patients with other neurological conditions (52% of MS vs 7% of non-MS in a multicenter study of 438 individuals)[26]. PRL have the promise of becoming a clinically relevant biomarker to improve both MS diagnosis and monitor treatment efficacy over time.

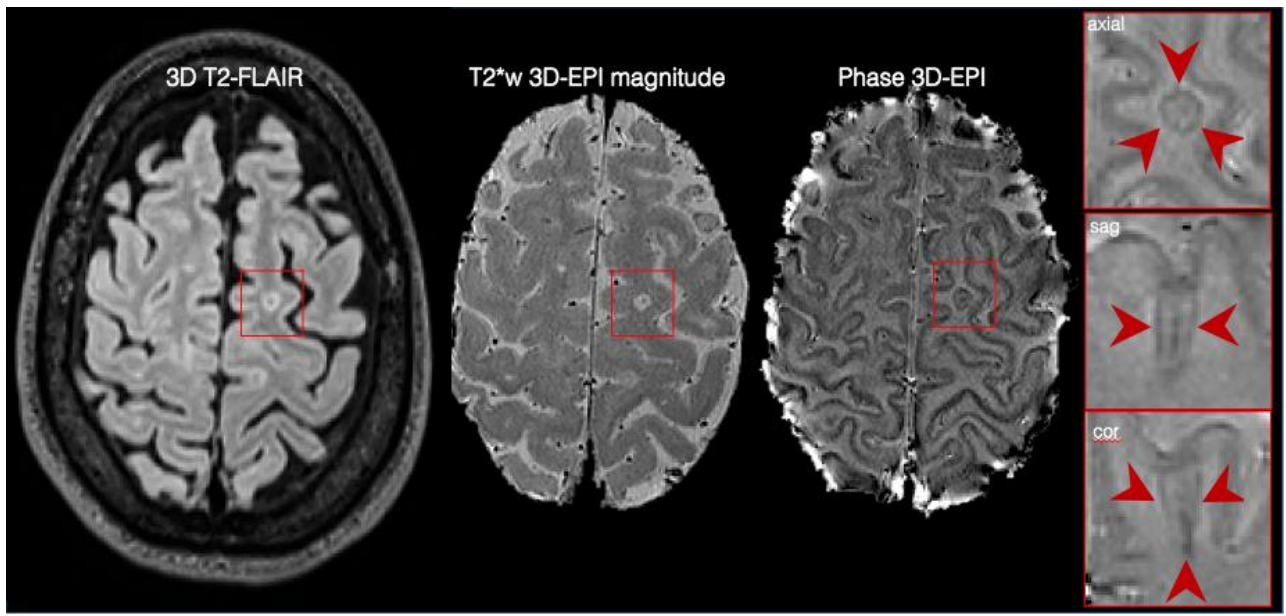

*Figure 3. A paramagnetic rim lesion seen on a 3T T2\*w 3D-EPI magnitude and phase 3D-EPI image in the three orthogonal planes (zoomed-in boxes). The CVS is also visible in the lesion.*

Overall, all three biomarkers have the potential to considerably improve the specificity of MS diagnosis[36,44,44]. Moreover, studies have shown that both CL and PRL can be useful to assess prognosis[30,61]. Their manual assessment, however, particularly for CL, is both time-consuming and prone to inter-rater variability. As for conventional WML, some automated or semi-automated methods have been proposed to accelerate this task. In the next section, we describe the challenges these approaches have been facing and how these differ from the segmentation of WML.

## 2.1 New challenges for machine learning methods

From a ML perspective, the automated segmentation or classification of advanced MS MRI biomarkers faces additional challenges as compared to their WML counterparts.

**Imaging guidelines-** The first obstacle is represented by the lack of consensus guidelines for imaging protocols. Although efforts have been made to standardize the use of MRI in clinical practice for conventional biomarkers[5], guidelines are still in a preliminary stage for advanced biomarkers. CL were included in the MS diagnostic criteria in 2017[3], but, currently, there is no single gold standard sequence at 3T for their detection in a clinical setting. PSIR, DIR, and MP2RAGE are all recommended by an international consensus[6]. However, these contrasts remain primarily acquired in research settings and are not yet widely used in clinical routine. Regarding the CVS, in 2016 a consensus statement by the North American Imaging in MS Cooperative (NAIMS) has proposed a standard radiological definition and suggested specific MRI acquisitions[24]. Following these recommendations, recent studies have shown that high-resolution T2*w 3D-EPI or FLAIR* improve the detection of the CVS compared to clinical acquisitions[51,52]. Nevertheless, a standardized clinical protocol for CVS detection is still missing. Among all advanced biomarkers, PRL is probably the one at the earliest stages. Although recent studies support the feasibility of its assessment on clinical scans and its utility in improving the diagnosis and prognosis of MS[26], there are not yet international guidelines for its definition nor a standardized MRI protocol for its analysis. Several different imaging modalities have been used for the PRL assessment, including phase 3D-EPI, susceptibility weighted imaging (SWI), QSM, and multi-echo T2* GRE at both 3T and 7T[60,61]. However, there is a paucity of studies that have systematically compared the sensitivity of these acquisition techniques for PRL detection, especially when implemented at different field strengths.

These evolving or unclear criteria for CL, the CVS and PRL, wide variety of imaging settings, and lack of clear guidelines for standardized protocols clearly jeopardize the development and wide use of these biomarkers and targeted ML techniques.

**Limited datasets -** An additional limitation, particularly for supervised DL-based approaches, is the scarcity and limited size of datasets in which these biomarkers were manually annotated. CL, CVS, and PRL all require for their assessment advanced MRI sequences at high or ultra-high magnetic field and experienced raters, and this makes it difficult to have large multi-site datasets. Although national MS registries exist in most countries, the data sharing of MRI in MS is still limited and often includes only conventional sequences[10]. Moreover, the CVS or the rim-shape in PRL are visible only on a few slices per lesion, reducing, even more, the data available to train a supervised approach. The presence of motion or other MRI artifacts can also result in the exclusion of poor quality scans reducing the data available for training/testing even further.

**Expert annotations -** Even for experts, the task of segmenting CL, detecting the CVS, or classifying PRL is intrinsically more challenging than segmenting WML. CL are generally smaller in size and more affected by partial volume (PV) effects, compared to WML. The cortex is convoluted, so lesion shape is not as regular as in WM, and traditional methods of radiological evaluation (scrolling through an image stack) are less effective in this context. The detection of the CVS requires susceptibility-based MRI and its exclusion criteria need to be carefully considered when performing its assessment[24]. T2*w phase images used to detect PRL present variability in the susceptibility signal and several artifacts, therefore experienced raters are needed. Moreover, as these three biomarkers have been so far mainly studied in research settings, clinicians do not commonly see them in clinical practice and might need specific training and dedicated time to perform a proper assessment.

**Inter-rater variability -** The lack of standardization for the definition and imaging of these biomarkers combined with the difficulty of the task, contributes to a modest inter-rater variability. Barquero et al.[66] showed that, in a cohort of 124 MS patients, approximately 38% of PRL needed a consensus review from two raters classifying PRL independently (Kappa score of 0.73). Absinta et al. observed similar inter-rater agreement between three experts at 3T (Fleiss coefficient of 0.71), with somewhat higher intra-rater reliability (Cohen k of 0.77)[60]. Similar values were reported at 7T for the same set of patients, whereas the agreement between 3T and 7T annotations was substantial (Cohen k of 0.78). In a similar way, the inter-rater agreement was shown to be moderate for the segmentation of CL[28,67] and moderate to high for the CVS[68,69]. Imaging quality and motion artifacts are other factors to consider as they can result in inconspicuity of all three biomarkers and, therefore, contribute to poor inter-rater agreement. Overall, the inter-rater variability represents an additional challenge for the development of automated approaches, as there might be large inconsistencies in the annotations of the training or testing set due to different raters performing the manual assessment.

## 3 Methods

Despite the recent discovery of the CVS and PRL and the above-mentioned challenges, few groups have already attempted to support their analysis with automated or semi-automated methods. To these two novel biomarkers, we add also CL, which, although studied for several years, have only recently been assessed automatically. Overall, much fewer methods have been proposed for the assessment of these three advanced biomarkers compared to WML (see Figure 5). In what follows, we briefly describe these state-of-the-art techniques by grouping them according to the biomarker they assess. A summary of the main characteristics for each method is presented in Table 1.

| Biomarker | Authors (year) | Method | Task | MRI sequences (magnetic field strength) | Dataset size (n. of sites) | Code available |
|---|---|---|---|---|---|---|
| **Cortical lesions** | Tardif, C. L., et al.[70] (2010) | Laminar profile shape analysis | S | Quantitative high-resolution scan (3T) | 1 post mortem brain scan (1) | No |
| | Fartaria, M.J., et al.[71] (2016) | K-NN | S | FLAIR, MPRAGE, MP2RAGE, DIR (3T) | 39 MS patients (1) | No |

| | Fartaria, M.J., et al.[72] (2017) | K-NN with partial volume constraints | S | FLAIR, MPRAGE, MP2RAGE, DIR (3T) | 39 MS patients (1) | No |
|---|---|---|---|---|---|---|
| | Fartaria, M.J., et al.[73] (2019) | PV estimation and topological constraints | S | MP2RAGE (7T) | 25 MS patients (2) | No |
| | La Rosa, F., et al.[74] (2020) | CNN | S | FLAIR, MP2RAGE (3T) | 90 MS patients (2) | Yes[a] |
| | La Rosa, F., et al.[75] (2020) | CNN | D and C | MP2RAGE, 2D T2*-w GRE, T2*w 3D-EPI (7T) | 60 MS patients (1) | Yes[b] |
| **Paramagnetic rim lesions** | Barquero, G., et al.[66] (2020) | CNN | C | FLAIR, T2*w 3D-EPI (3T) | 124 MS patients (2) | No |
| | Lou, C., et al.[76] (2021) | Random forest classifier | C | FLAIR, T1-w, T2*w 3D-EPI (3T) | 20 MS patients (1) | Yes[c] |
| **Central vein sign** | Maggi, P., Fartaria, MJ, et al.[69] (2020) | CNN | C | T2*w 3D-EPI, FLAIR (3T) | 42 MS patients, 33 mimics, 5 others (3) | No |
| | Dworkin, J. D., et al.[77] (2018) | Probabilistic method | C | T2*w 3D-EPI, FLAIR (3T) | 16 MS patients, 15 MS mimics (1) | Yes[d] |

*Table 1. Summary of the methods proposed for the automated or semi-automated analysis of cortical lesions, the central vein sign, and paramagnetic rim lesions. The task is abbreviated as*

---

[a] https://github.com/FrancescoLR/MS-lesion-segmentation
[b] https://github.com/Medical-Image-Analysis-Laboratory
[c] https://github.com/carolynlou/prlr
[d] https://github.com/jdwor/cvs

follows: segmentation (S), classification (C), and detection (D). If not specified, all sequences were 3D.

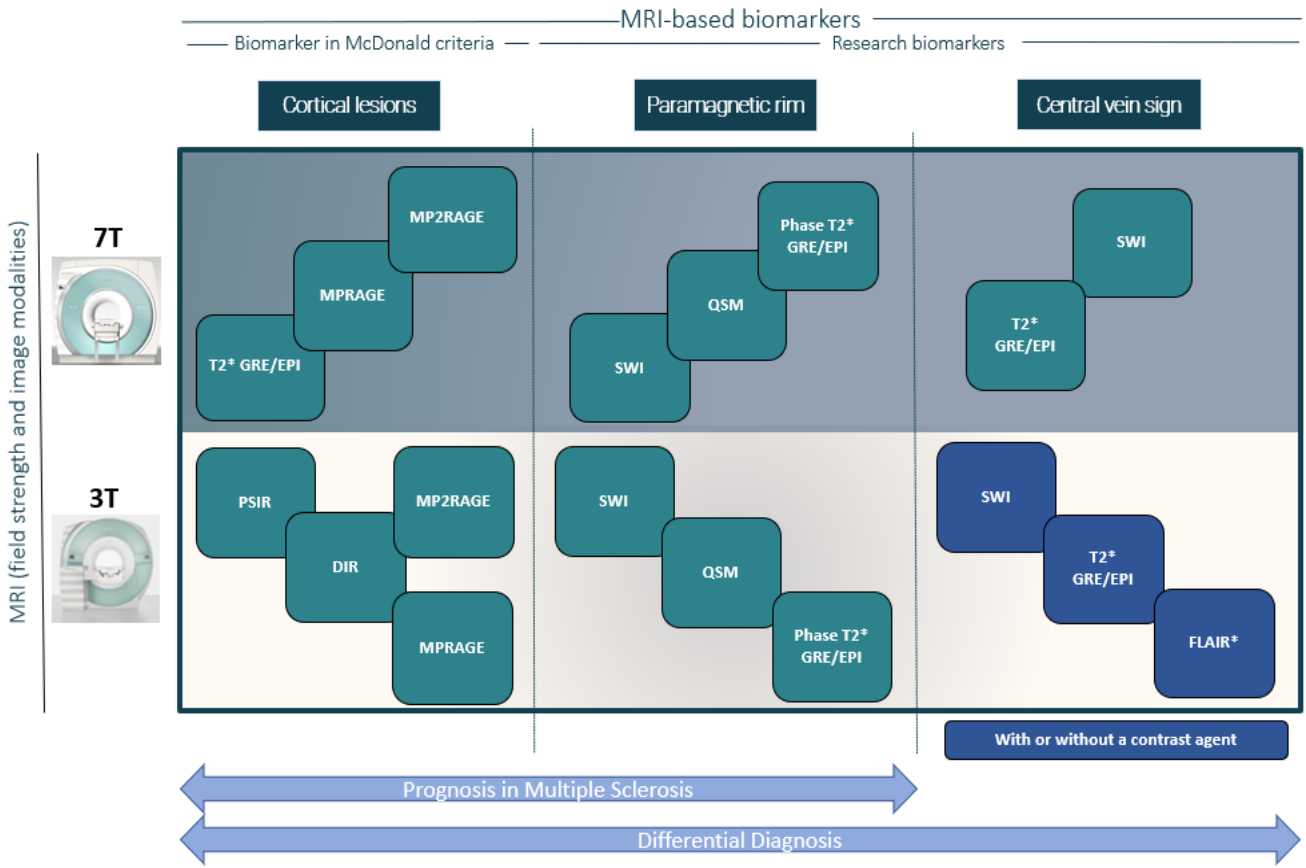

Figure 4. Scheme showing the main MRI sequences used for detecting each biomarker at both 3T and 7T.

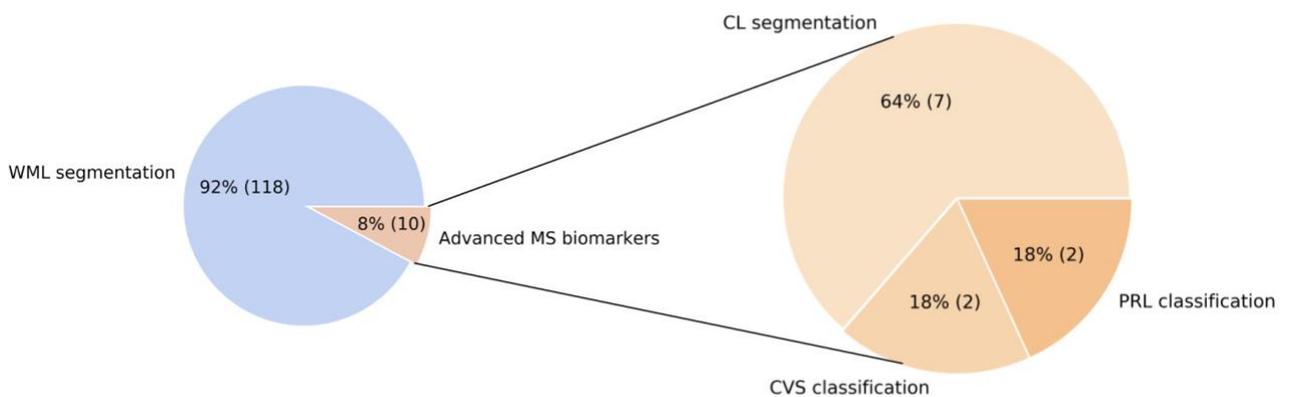

Figure 5. Pie charts showing the approximate proportion of automated methods proposed for WML segmentation and advanced biomarkers assessment (left) and the proportions for the single advanced biomarkers (right). The total number of automated WML segmentation methods (118) was obtained by adding the studies analyzed in the following surveys[14,78,79]. The advanced MS biomarker methods were retrieved from the literature.

## 3.1 Cortical lesions

ML-based methods automatically segmenting CL have been explored with both 3T and 7T MRI. The first work present in the literature considered a postmortem MS brain imaged at 3T with high resolution (0.35mm isotropic)[70]. Tardif et al. proposed to first identify the cortical and white matter surfaces, then extract laminar profiles between the two tissues, and finally apply a k-means classifier to the profile intensity and shape features to parcellate the cortex and detect lesions. Although showing promising results on one postmortem MS brain, this method was never validated with larger cohorts nor in-vivo data. A few years later, Fartaria et al.[71] proposed the first automated method for the segmentation of both WM and cortical lesions. In their study, they analyzed a cohort of 39 early-stage MS patients and considered both conventional (FLAIR, MPRAGE) and advanced (MP2RAGE, DIR) MRI sequences at 3T. In a nutshell, their method consisted of co-registering the different MRI contrasts, leveraging prior tissue probability maps from existing brain atlases of healthy subjects, and finally classifying each voxel either as being a lesion or healthy tissue with a k-nearest neighbor (k-NN) algorithm. Additionally, as post-processing, lesions smaller than 3.6 µL were discarded, and a region-growing algorithm was applied to improve the lesion delineation. Results were promising, showing a CL detection rate of 62% when advanced imaging (FLAIR, MP2RAGE, and DIR) was included. An extension of this segmentation framework with a Bayesian partial volume (PV) estimation method was presented by the same authors[72]. They argued that CL, being generally small and located at the interface between WM and GM, suffer from strong PV effects. The addition of this PV model indeed improved the delineation of CL, while the detection rate remained similar (58%)[72].

The same research group also proposed a different segmentation method for WML and CL using only 7T MP2RAGE images (called MSLAST: Multiple Sclerosis Lesion Analysis at Seven Tesla)[73]. MSLAST computes tissue concentration maps with a PV algorithm and unifies them based on topological constraints. A connected-components analysis is then performed on gray matter and cerebrospinal fluid maps, and small components are classified as MS lesions. This method was evaluated with 25 MS patients' scans from two research centers and reached a 58% patient-wise CL detection rate (when 6µL was considered as minimum lesion volume) with a false positive rate of 40%. Moreover, it showed promising scan-rescan repeatability within the same session, with a mean total lesion volume difference (WML and CL combined) of 0.29 mL (mean total lesion volume 5.52 mL), vs 0.13 mL for the manual segmentations. More recently, DL-based approaches have been presented as well[74,75]. In the first study, La Rosa et al. proposed a framework for the automated segmentation of WML and CL at 3T using FLAIR and MP2RAGE[74]. Their method extracts 3D patches of 88x88x88 voxels from the two MRI contrasts and feeds them to a convolutional neural network (CNN). The CNN, inspired by the U-Net, has an encoder and decoder path, each one with three resolution levels. Evaluated on two datasets acquired in different centers, for a total of 90 MS patients, the framework showed competitive performance, with a CL detection rate of 76% and a false positive rate of 29%. In a second study, the same group proposed a similar approach, this time tailored exclusively for the detection of CL using multi-contrast 7T MRI[75]. The contrasts considered were MP2RAGE, T2*-weighted GRE, and T2*-w 3D-EPI. A cohort of 60 patients was analyzed with a total of over 2000 CL manually segmented by two experts. The CNN architecture proposed was similar to the one just described, but with a modified output. In addition to the CL segmentation, the CNN provided a classification into two types (leukocortical and intracortical/subpial lesions) and a separate branch with a simple tissue segmentation in WM/GM. CL were correctly classified into the two types by the network with an accuracy of 86%. Setting a minimum lesion size of 0.75 µL, it achieved a CL detection rate of 67% with, however, a quite high false positive rate of 42%. Importantly, about 24% of these false positives were retrospectively judged as CL or possible CL by an expert[75].

## 3.2 The central vein sign

The automated and standardized detection of the central vein sign could potentially be of clinical utility in improving the accuracy of MS diagnosis. As of today, two automated methods for the classification of MS lesions as CVS+ (MS lesions showing the presence of the CVS) or CVS- (MS lesions without the CVS) have been proposed in the literature[69,77]. Dworkin et al.[77] proposed a probabilistic method based on the Frangi vesselness filter[80]. They first perform an automated WML segmentation using T1 and FLAIR 3D MRI volumes acquired at 3T[81,82] and obtain a map of the veins by applying the vesselness filter to a T2*w 3D EPI image. Confluent lesions are then separated and lesion centers are detected by textural analysis[83]. Periventricular lesions are discarded as suggested by consensus guidelines[24], and a permutation algorithm is applied to verify whether veins occur at the lesions' centers more often than would be expected due to random chance. Finally, to account for scan motion, the single lesion CVS+ probabilities are weighted by the noise in their T2*-w 3D-EPI intensities and averaging across the total number of lesions, a patient-wise CVS value is obtained. This method was evaluated on a cohort of 31 adults, of whom 16 had MS. When considering a 40% cutoff rule, the method yielded a sensitivity of 0.94 and a specificity of 0.67 on a patient-wise classification level. The performance of the method on a lesion-wise level was not assessed. Although still far from experts' performance, this was a first attempt to automatize the CVS assessment and encouraged further improvements.

Maggi, Fartaria et al.[69] introduced an optimized CNN for the automated CVS assessment, called CVSnet. CVSnet is inspired by the VGGnet[84] but composed of only three convolutional layers followed by rectified linear unit (ReLU) activations. Dropout was applied in each layer, and then two fully-connected layers of size 32 and 2, respectively, were added to provide the output. The authors selected 3D patches of size 21x21x21 voxels as input for the network, where each patch was centered on an MS lesion and FLAIR* was the only MRI contrast used. Moreover, an ensemble of 10 networks with the same architecture was trained and the probability outputs were averaged to provide the final prediction. This study considered a cohort of 80 patients imaged at three different sites, of whom 42 had MS, 35 an MS-mimic, and 5 an unknown diagnosis. On the test set, CVSnet significantly outperformed the vesselness filter[80] (p<0.001), reaching a sensitivity, specificity, and accuracy of 0.83, 0.75, and 0.79, respectively, approaching expert performance. However, as argued by the authors, these results are not directly comparable with those of Dworkin at al.[77], as the CVSnet considered different exclusion criteria to pre/select the lesions, and the initial lesion segmentation was performed manually.

### 3.3 Paramagnetic rim lesions

To our knowledge, only two methods have been proposed so far for the detection of rim-like features and classification of PRL[66,76]. Barquero et al. introduced a DL-based approach (called RimNet) for the semi-automated classification of PRL, which considered 3D FLAIR and T2*w 3D-EPI and phase 3D-EPI images. RimNet's architecture is inspired by the VGGnet[84] and composed of two parallel CNN (one for either FLAIR or T2*w 3D-EPI image and one for the phase 3D-EPI image), where each CNN is made of three convolutional layers followed by a max-pooling operation. 3D patches of size 28x28x28 (centered around each MS lesion) are fed to each branch, and both high-level and low-level feature maps are concatenated. An automated lesion segmentation based on FLAIR and

MPRAGE/MP2RAGE[74,85] was modified by an expert to split confluent lesions. The performance of RimNet was assessed on a cohort of 124 adults with MS who underwent 3T MRI at two different sites with two scanners from the same vendor. Two experts annotated PRL independently and reached consensus in a joint session (462 PRL in total). The proposed multimodal approach based on FLAIR and phase 3D-EPI image achieves lesion-wise sensitivity and specificity of 0.70 and 0.95, respectively. When considering a previously identified clinical threshold of 4 PRL[22] for classifying patients as "chronic active" and "non-chronic active", RimNet reaches an accuracy of 0.90 and an F1-score of 0.84. These values are within 5% of the single experts' metrics, suggesting that RimNet could be a valuable tool in supporting the PRL analysis. The main drawback of RimNet, however, is that the method is not fully automated, as confluent lesions were split manually by an expert.

Lou et al.[76], on the other hand, proposed a fully automated method for PRL assessment. They considered a cohort of 20 subjects with MS imaged with 3D FLAIR, 3D MPRAGE, and T2*-w 3D-EPI and phase 3D-EPI images. One neurologist inspected the T2* magnitude and unwrapped phase images and annotated PRL (113 PRL over the entire cohort). The automated pipeline, after some pre-processing steps that included lesion segmentation[81,82], lesion center detection[83], and lesion labeling, consisted of extracting 44 different lesion-wise radiomic features. A random forest classifier was then fitted on these features, and its ability to classify PRL was evaluated on a test set of 4 patients. Sensitivity and specificity of 0.75 and 0.81, respectively, were achieved. Although fully automated, this study has three limitations. First, the extremely small testing dataset (4 patients only with 47 PRL), annotated by a single expert, does not guarantee the generalization of the proposed method. Second, all patients analyzed had at least one PRL, and this might add a bias to the trained model. Finally, as acknowledged by the authors, about 65% of misclassified lesions were confluent, highlighting the need for a better solution to address these lesions.

## 4  Discussion

The methods described here tackle challenging and clinically relevant problems. Automated and reliable solutions for detecting, classifying, and segmenting CL, CVS PRL are needed to improve the standardization of these biomarkers and facilitate their assessment in clinical routine. As of today, however, these methods are still in an early stage and are slightly less sensitive than WML segmentation approaches.

Nevertheless, such tools would provide obvious advantages, either as stand-alone or adjunctive approaches, particularly for features (like PRL and CVS) that are difficult and time-consuming to analyze using conventional radiological workflows - in these particular cases, manual reading is so involved, that automated methods might actually boostr widespread adoption. First, they can substantially reduce analysis time, as compared to a manual rating. Maggi, Fartaria et al., for instance, showed that the automated CVS assessment proposed was 600-fold faster on the test set compared to the manual assessment (4 seconds vs 40 minutes)[69]. Similar values were reported also for CL, whereas they were not specified for the PRL analysis. A second main advantage of automated methods is their ability to base their decision on 3D multi-contrast MRI analyzed simultaneously. This stands in contrast to expert reviews, which typically involve comparison of 2D slices across several contrast mechanisms in a variety of planes and are thus inherently limited in the amount of information that can be readily gleaned.

## 4.1 Common trends

Some common trends can be observed in most of the proposed pipelines. The large majority of the methods are supervised, relying on expert annotations. Regarding the DL-based approaches, they all used patch-based 3D CNN, exploiting the 3D intrinsic information, and often considered more than one MRI contrast simultaneously. In addition, a shared tendency consists in the use of relatively shallow architectures, with a limited number of trainable parameters, due to the lack of large datasets[66,69,74,75]. Combining this with extensive data augmentation techniques can help when datasets are small and unbalanced. Alternatively, other groups have tackled the problem of overfitting by proposing approaches based on classical ML techniques, such as k-NN[71,72] or random forest classifier[12]. In these studies, either intensity-based, radiomic, or probabilistic features are extracted and then fed to the respective classifier. Overall, their current performance is inferior compared to their DL-based counterparts.

In addition, some common pre-processing steps can also be identified. First, some methods use intensity normalization techniques, either based on the entire 3D volumes[71,72,74–77] or on single patches[66,69]. Second, the approaches using multiple MRI contrasts always register all images to the same space[71,72,74–76]. Registration errors might affect the methods' performance. Finally, a shared pre-processing step in all approaches for the CVS or PRL assessment is the prior WML segmentation, obtained either manually[69] or with an automated tool[66,76,77]. In both cases, this can be a source of error that negatively affects the subsequent biomarkers' classification accuracy.

## 4.2 Current limitations

Currently, a major limitation hinders the deployment of the above-described methods to the clinic: the methods proposed were trained and evaluated on small datasets acquired from one or at most two centers. Moreover, the MRI protocols used were often similar and not representative of the current diversity of images acquired in the clinics, including scans affected by noise and artifacts or protocols missing certain modalities. Therefore, the automated methods' robustness on larger datasets and different scanners, especially from multiple vendors, remains to be proven. This limitation is emphasized by the current lack of standardized acquisition protocols which increases the diversity of the MRI sequences considered for the same biomarkers. This represents also a major hurdle for potential regulatory approval of such methods. As regulatory approval is necessary for widespread adoption in the clinics, which is, in turn, the prerequisite for the availability of large datasets, this is almost a catch-22.

In addition, the achieved performance levels of the automated methods are still inferior compared to the human experts. Considering the high inter-rater variability and the limited amount of data available, there is also a considerable risk of having methods that perform well on data annotated by a single expert and not as well with annotations from other raters. To mitigate this issue, several methods have already considered consensus annotations from two or more experts[66,69,75]. Regarding CL, no automated method presented in the literature was compared, on the same dataset, with the experts' inter-rater variability, thus a proper evaluation is not possible. With respect to CVS, Maggi, Fartaria et al.[69] compared the performance of CVSnet with the consensus of two experts. Following the "50% rule," CVSnet achieved on the testing set a classification accuracy of 79%, whereas the experts reached 100% accuracy in differentiating MS and mimic diseases. In a similar way, Barquero et al.[66] compared RimNet's performance with those of two experts in classifying PRL. In a lesion-wise analysis, RimNet achieved a sensitivity of 71% and a negative predictive value of 96%, approaching the experts, who reached 78% and 98%, respectively.

Another main limitation is represented by the fact that some methods presented are not fully automated. CVSnet[69], for instance, used manually annotated MS lesion masks in which lesions were excluded based on the NAIMS criteria[24], whereas in the pipeline proposed by Dworkin et al.[77], scans affected by noise were discarded following a manual rating. Similarly, RimNet[66] exploits lesion masks where confluent lesions have been previously split into single units by an expert. In contrast, all methods described to date for CL segmentation or detection are fully automated[71,72,74,75]. Another persistent issue in the automated analysis of the CVS and PRL is the presence of confluent lesions. Large, periventricular white matter lesions which include several single units pose additional challenges as the current methods classify each lesion singularly[76,77], and some of them extract 3D patches centered on the lesion of interest[66,69]. In RimNet[66], for instance, an expert manually split confluent lesions, whereas Lou et al. observed a consistent drop in performance in PRL classification in the presence of confluent lesions[76]. Although methods to automatically split confluent lesions have been proposed[83,86], further developments are needed in order to properly apply these in the presence of the CVS or PRL.

Finally, for every automated tool the regulatory environment remains a critical barrier, as up to date less than 90 AI/ML-based medical devices or algorithms have been approved by the US Food & Drugs Administration (FDA). This challenge, however, is not unique to advanced biomarkers[87] but shared also by automated approaches segmenting WML or estimating brain atrophy.

### 4.3  Future research avenues

**Standardization -** The first two necessary steps toward the improvement of the above-referred approaches are the validation of the biomarkers' specific criteria and standardization of the relative MRI protocols. CL have been recently included in the MS diagnostic criteria[3], however, a consensus on imaging and on their definition is still missing. In a similar way, PRL urgently need a consensus definition and standardized clinical protocols, whereas the initial criteria proposed for the CVS[24] need to be updated in light of the latest studies. This would clarify the automated methods' goals, which so far have been extremely dependent on specific expert labeling of each dataset, or the specific criteria adopted.

**Extensive validation and generalization -** Generalization of the proposed methods needs to be validated on large, multi-site datasets. For this purpose, we urge research groups to organize grand challenges and release publicly available datasets with advanced biomarkers. As already proved for several other tasks in medical imaging[88], as well as for WML segmentation[11,12], such open data initiatives boost on the one hand the development of new methods, and on the other hand, help set benchmarks for a fair assessment. Only 4 of the 10 methods covered in this review publicly released their code. To extend their usage and foster a culture of open science, research groups should make their code publicly available and possibly provide Docker/Singularity[e] images to facilitate their evaluation. Moreover, as has been successfully done for WML segmentation[89], domain-adaptation techniques should also be explored in order to improve robustness of the automated methods to noise, artifacts, and different protocols. So far, all three biomarkers have been primarily studied at 3T and 7T, and therefore robust methods able to work with images acquired at both magnetic field strengths would be very valuable.

**Transfer learning -** Considering the scarcity of large, annotated datasets, an additional strategy that should be explored consists of transfer learning. Sharing of neural network weights between research groups could foster interdisciplinary applicability of CNN trained on relatively large datasets towards different purposes, such as advanced MS biomarkers, by fine-tuning the trained models in smaller datasets. Potential advantages would include a shorter training time and robust feature extraction across different MRI device manufacturers or different pulse sequence acquisition parameters[90].

**Longitudinal assessment -** Another possible research direction is an expansion of the current methods to analyze longitudinal data. To the best of our knowledge, only one study has tackled the longitudinal assessment of CL at 3T[91], whereas PRL evolution over time has not yet been assessed with automated approaches. CL are known to play a major role in disease progression[32] and considerable changes in their volume were observed in longitudinal studies[92,93]. Of similar interest, PRL and slowly-expanding lesions (SELs) volume assessment over time is a plausible future clinical measure of treatment response[64,94–96]. Therefore, automated longitudinal assessment of both CL and PRL could be of high relevance.

---

[e] https://www.docker.com/, https://sylabs.io/singularity

**Joint assessment of multiple biomarkers-** To date, all the methods proposed tackled the assessment of a single advanced biomarker, although in the case of CL some methods consider WML as well[73,74,97]. Future work may aim at automatically analyzing multiple biomarkers in a unified framework (eg. with the same input images and algorithm) as this would be extremely useful for research purposes or in clinics. Moreover, ML-based algorithms have the potential to be useful also for prediction purposes. A few automated methods based either on MRI[98–100], optical coherence tomography[101], or clinical information[87] have already been presented to predict MS progression. Specifically to advanced biomarkers, Traeba et al. have proposed a ML approach for the regression of both CL and PRL, in the same cohort of patients, with disability progression[102]. In this prospective, longitudinal study, the authors analyzed brain scans of 100 MS patients using 7T susceptibility-sensitive MRI in which CL and PRL were segmented manually. Although the study had some limitations, including the fact that the disability progression was assessed only by the EDSS and only one ML-based method (gradient boosting algorithm, XGBoost) was tested, it showed that 7T MRI and the combination of different biomarkers are promising in predicting MS disability progression. Future studies should aim to combine the automated assessment of multiple biomarkers with clinical information and other relevant markers to predict clinical outcomes or treatment effect.

**Explainable AI -** As discussed in this paper, ML methods combined with specialized MRI sequences could play a fundamental role in supporting the diagnosis of, and prognostication in, MS. However, the complexity of DL algorithms hinders their interpretation, which has led some to consider these methods as "black boxes." The lack of an obvious connection between biology, pathophysiology, and features revealed by DL might diminish clinicians' confidence in these algorithms, again hindering the adoption of such tools in clinical research and healthcare. Explainable AI (XAI) methods are needed as to on one side provide uncertainty and on the other side transparency on the decisions taken by the DL-models. By explainability, we refer to a set of domain features such as pixels of an image or human-understandable high-level attributes that contribute to the output decision of the model and its internal working. To our knowledge, there are only two groups that have investigated XAI in MS. Eitel et al.[103] explore explainability to reveal relevant voxel-wise locations that a trained CNN uses for distinguishing between a normal and MS brain MRI. They found that diagnostic success relied on the appearance of both lesions and non-lesional tissue (thalamus). Nair et al.[104] studied the uncertainty of DL-based lesion segmentation to quantify the AI model reliability. Interestingly, their results showed that discarding lesions with high estimated uncertainty from the output segmentation would improve the performance of the model. These two pioneering approaches strengthen the idea that explainability and uncertainty measures can reliably provide new insights into how DL models for MS work and potentially improve them and increase their transparency.

Overall, we believe that developing explainable AI tools is crucial in the ML MS research roadmap and would have an impact at both methodological and clinical levels. First, explainable DL in MS would provide new insights into model decisions and help identify any bias. Second, the inclusion of uncertainty and explainability will help in increasing the confidence of clinicians considering their use, as well as improve the quality of decision making and ultimately the clinical impact. Finally, they may foster a better understanding of MS progression by generating biologically interpretable measures of inflammation and degeneration.

### 4.4 Conclusion

To summarize, automated or semi-automated methods aiming to segment and classify CL, CVS, and PRL are still in an early stage. Nevertheless, these pioneering methods have the potential to improve the biomarkers' standardization and facilitate their large-scale assessment in clinical routine. Automated or semi-automated tools could considerably reduce the current amount of time needed for a manual assessment. To date, however, some limitations still hinder the broader deployment of these tools. A major barrier is their lack of validation on multi-center datasets acquired with different protocols. To overcome this limitation there is an urgent need for consensus criteria and standardized clinical protocols for all three advanced biomarkers. Future work should focus on improving the robustness of the automated methods, extending their framework with longitudinal data, and including interpretable measures into their decisions. Finally, we encourage research groups to organize grand challenges and release publicly available datasets. This would boost the development of new methods and provide benchmarks for a fair comparison that is currently missing.

**Acknowledgments**

We acknowledge access to the facilities and expertise of the CIBM Center for Biomedical Imaging, a Swiss research center of excellence founded and supported by Lausanne University Hospital (CHUV), University of Lausanne (UNIL), Ecole Polytechnique fedérale de Lausanne (EPFL), University of Geneva (UNIGE) and Geneva University Hospitals (HUG). We thank Thomas Yu for proofreading the manuscript.

**Conflict of interest**

Till Huelnhagen and Tobias Kober are employed by Siemens Healthcare AG Switzerland. Cristina Granziera has received the following fees which were used exclusively for research support: (i) advisory board and consultancy fees from Actelion, Genzyme-Sanofi, Novartis, GeNeuro, and Roche; (ii) speaker fees from Genzyme-Sanofi, Novartis, GeNeuro, and Roche; (iii) research support from Siemens, GeNeuro, Roche. Daniel Reich received research support from Abata, Sanofi-Genzyme, and Vertex. Pietro Maggi received support from Biogen and Cliniques universitaires Saint-Luc Fonds de Recherche Clinique. Russell Shinohara receives consulting income from Octave Bioscience, and compensation for reviewing scientific articles from the American Medical Association. Martina Absinta received consultancy fees from GSK and Sanofi-Genzyme. The other authors have nothing to declare.

**Funding**

F.L.R. is supported by the European Union's Horizon 2020 research and innovation program under the Marie Sklodowska-Curie project TRABIT (agreement No 765148) and the Novartis Research Foundation. M.W. is supported by the Swiss Government Excellence Scholarship for Foreign Scholars (No 2021.0087). O.A. is supported by a National Multiple Sclerosis Society (NMSS) - American Brain Foundation Clinician Scientist Development Award (FAN-1807-32163). E.S.B. is supported by a Career Transition Fellowship from the National Multiple Sclerosis Society. P.S., O.A., E.S.B., and D.S.R. are supported by the Intramural Research Program of the National Institute of Neurological Disorders and Stroke, National Institutes of Health, Bethesda, Maryland, USA. R.T.S. is partially supported by R01NS060910, R01MH112847, R01MH123550, U01NS116776, and R01NS112274 from the National Institutes of Health. The content is solely the responsibility of the authors and does not necessarily represent the official views of the National Institutes of Health. M.A. is supported by the Conrad N. Hilton Foundation (Marilyn Hilton Bridging Award for Physician-

Scientists, grant #17313), the International Progressive MS alliance, the Roche Foundation for Independent Research, the Cariplo Foundation (grant #1677), and the FRRB Early Career Award (grant#1750327).